\documentclass{article}
\usepackage[utf8]{inputenc}

\usepackage{graphicx}
\usepackage{multirow}
\usepackage[margin=1.5in]{geometry}
\usepackage{setspace}
\usepackage{amsmath}
\usepackage{mathpazo}
\usepackage{marginnote}
\usepackage{multirow}
\usepackage{fancyhdr}
\usepackage{lastpage}
\usepackage{url}
\usepackage{pgfplots,tikz}

\title{NIST SRE CTS Superset: A large-scale dataset for telephony speaker recognition}
\author{Omid Sadjadi}
\date{August 2021}

\graphicspath{{./figs/}}

\begin{document}

\maketitle

\section{Introduction}
This document provides a brief description of the National Institute of Standards and Technology (NIST) speaker recognition evaluation (SRE) \cite{greenberg2020two} conversational telephone speech (CTS) Superset. The CTS Superset has been created in an attempt to provide the research community with a large-scale dataset along with uniform metadata that can be used to effectively train and develop telephony (narrowband) speaker recognition systems. It contains a large number of telephony speech segments from more than 6800 speakers with speech durations distributed uniformly in the [10s, 60s] range. The segments have been extracted from the source corpora used to compile prior SRE datasets (SRE1996-2012), including the Greybeard corpus as well as the Switchboard and Mixer series collected by the Linguistic Data Consortium (LDC\footnote{https://www.ldc.upenn.edu/}). In addition to the brief description, we also report speaker recognition results on the NIST 2020 CTS Speaker Recognition Challenge, obtained using a system trained with the CTS Superset. The results will serve as a reference baseline for the challenge.

\section{CTS Superset (LDC2021E08)}
The NIST SRE CTS Superset is the largest most comprehensive dataset available to date for telephony speaker recognition. It has been extracted from the source corpora (see Table~\ref{tab:source}) used to compile prior SRE datasets (SRE1996-2012). Table~\ref{tab:stats} summarizes the data statistics for the CTS Superset. There are a total of 605,760 segments originating from 6867 speakers (2885 male and 3992 female). Each segment contains approximately 10 to 60 seconds of speech \footnote{As determined using a speech activity detector (SAD).}, and each speaker has at least three sessions/calls (hence at least 3 segments). Note that some speakers appear in more than one source corpus, therefore the sum of the speakers in the table is greater than 6867. Although the vast majority of the segments in the CTS Superset are spoken in English (including both native and accented English), there are more than 50 languages represented in this dataset.

The procedure for extracting segments from the original sessions/calls is as follows; given a session of arbitrary duration (typically 5--12 minutes) and speech time marks generated using a speech activity detector (SAD), we start extracting non-overlapping segments by repeatedly sampling a speech duration from the uniform distribution [10, 60], until we exhaust the duration of that session.

\begin{table}[t]
    \centering
    \caption{Original source corpora used to create the NIST SRE CTS Superset}
    \begin{tabular}{|l|c|c|}
    \hline
         \textbf{Source corpus} & \textbf{LDC Catalog ID} & \textbf{corpusid} \\ \hline\hline
         Switchboard1 release2 & LDC97S62 \cite{swb1} & swb1r2\\ \hline 
         Switchboard2 Phase I & LDC98S75 \cite{swb2p1} & swb2p1 \\ \hline
         Switchboard2 Phase II & LDC99S79 \cite{swb2p2} & swb2p2 \\ \hline
         Switchboard2 Phase III & LDC2002S06 \cite{swb2p3} & swb2p3  \\ \hline
         Switchboard Cellular Part 1 & LDC2001S13 \cite{swbcellp1} & swbcellp1 \\ \hline
         Switchboard Cellular Part2 & LDC2004S07 \cite{swbcellp2} & swbcellp2 \\ \hline
         Mixer3 & LDC2021R03 \cite{ldc2007} & mx3 \\ \hline
         Mixer4--5 & LDC2020S03 \cite{mx45} & mx45\\ \hline
         Mixer6 & LDC 2013S03 \cite{mx6} & mx6 \\ \hline
         Greybeard & LDC2013S05 \cite{gb1} & gb1\\ \hline
    \end{tabular}
    
    \label{tab:source}
\end{table}

\begin{table}[t]
    \centering
    \caption{Data statistics for the NIST SRE CTS Superset}
    \begin{tabular}{|l|c|c|c|}
    \hline
         \textbf{corpusid} & \textbf{\#segments} & \textbf{\#speakers} & \textbf{\#sessions}  \\ \hline\hline
         swb1r2 & 26,282 & 442 & 4757 \\ \hline 
         swb2p1 & 33,746 & 566 & 7134 \\ \hline
         swb2p2 & 41,982 & 649 & 8895 \\ \hline
         swb2p3 & 22,865 & 548 & 5187 \\ \hline
         swbcellp1 & 13,496 & 216 & 2560 \\ \hline
         swbcellp2 & 20985 & 378 & 3966 \\ \hline
         mx3 & 317,950 & 3033 & 37,759 \\ \hline
         mx45 & 40,313 & 486 & 4997 \\ \hline
         mx6 & 70,174 & 526 & 8727 \\ \hline
         gb1 & 17,967 & 167 & 2188 \\ \hline
    \end{tabular}
    
    \label{tab:stats}
\end{table}

The LDC2021E08 package contains the following items:

\begin{itemize}
\item Audio segments from 6867 subjects located in the \texttt{data/\{subjectids\}/} directories
\item Associated metadata located in the \texttt{docs/} directory
\end{itemize}
The metadata file contains information about the audio segments and includes the following fields:
\begin{itemize}
    \item \texttt{filename} (segment filename including the relative path)
    \item \texttt{segmentid} (segment identifier)
    \item \texttt{subjectid} (LDC speaker id)
    \item \texttt{speakerid} (zero indexed numerical speaker id)
    \item \texttt{speech\_duration} (segment speech duration)
    \item \texttt{sessionid} (segment session/call identifier)
    \item \texttt{corpusid} (corpus identifier as defined in Table~\ref{tab:source})
    \item \texttt{phoneid} (anonymized phone number)
    \item \texttt{gender} (male or female)
    \item \texttt{language} (language spoken in the segment)
\end{itemize}

For future releases of the CTS Superset we plan to extend the source corpora to include Mixer 1, 2, and 7 as well. 

\section{Speaker Recognition System}

\begin{figure}[h]
\centering
\includegraphics[scale=0.55, clip, trim=70mm 80mm 50mm 50mm]{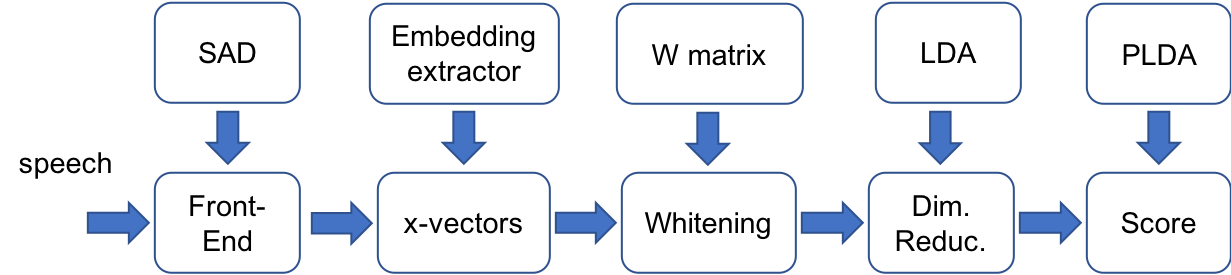}
\caption{Block diagram of the baseline system.}
\label{fig:blockdiag_xvec}
\end{figure}

In this section, we describe the baseline speaker recognition system setup including speech and non-speech data used for training the system components as well as the hyper-parameter configurations used. Figure~\ref{fig:blockdiag_xvec} shows a block diagram of the x-vector baseline system. The embedding extractor is trained using Pytorch\footnote{https://github.com/pytorch/pytorch}, while the NIST SLRE toolkit is used for front-end processing and back-end scoring.

\subsection{Data}

The baseline system is developed using the CTS Superset described in the previous section. In order to increase the diversity of the acoustic conditions in the training set, two different data augmentation strategies are adopted. The first strategy uses noise-degraded (using babble, general noise, and music) versions of the original recordings, while the second strategy uses spectro-temporal masking applied directly on spectrograms (aka spectrogram augmentation \cite{specaug}). The noise samples for the first augmentation approach are extracted from the MUSAN corpus \cite{musan}. For spectrogram augmentation, the mild and strong policies described in \cite{specaug} are used.

\subsection{Configuration}

For speech parameterization, we extract 64-dimensional log-mel spectrograms from 25 ms frames every 10 ms using a 64-channel mel-scale filterbank spanning the frequency range 80~Hz--3800~Hz. After dropping the non-speech frames using SAD, a short-time cepstral mean subtraction is applied over a 3-second sliding window.

For embedding extraction, an extended TDNN \cite{snyder2019} with 11 hidden layers and parametric rectified linear unit (PReLU) non-linearities is trained to discriminate among the nearly 6800 speakers in the CTS Superset set. A cosine loss with additive margin \cite{cosface} is used in the output layer (with $m=0.2$ and $s=40$). The first 9 hidden layers operate at frame-level, while the last 2 operate at segment-level. There is a 3000-dimensional statistics pooling layer between the frame-level and segment-level layers that accumulates all frame-level outputs from the 9\textsuperscript{th} layer and computes the mean and standard deviation over all frames for an input segment. The model is trained using Pytorch and the stochastic gradient descent (SGD) optimizer with momentum ($0.9$), an initial learning rate of $10^{-1}$, and a batch size of $512$. The learning rate remains constant for the first $5$ epochs, after which it is halved every other epoch. 

To train the network, a speaker-balanced sampling strategy is implemented where in each batch 512 unique speakers are selected, without replacement, from the pool of training speakers. Then, for each speaker, a random speech segment is selected from which a 400-frame (corresponding to 4 seconds) chunk is extracted for training. This process is repeated until the training samples are exhausted.

After training, embeddings are extracted from the 512-dimensional affine component of the 10\textsuperscript{th} layer (i.e., the first segment-level layer). Prior to dimensionality reduction through linear discriminant analysis (LDA) to 250, 512-dimensional embeddings are centered, whitened, and unit-length normalized. The centering and whitening statistics are computed using the CTS Superset data. For backend scoring, a Gaussian probabilistic LDA (PLDA) model with a full-rank Eigenvoice subspace is trained using the embeddings extracted from  only the original (as opposed to degraded) speech segments in the CTS Superset. No parameter/domain adaptation or score normalization/calibration is applied.

\section{Results}

In this section, we present the experimental results on the NIST 2020 CTS Challenge progress and test sets obtained using the baseline system. Results are reported in terms of the equal error rate (EER) and the minimum cost (denoted as min\_C) and defined in the CTS Challenge evaluation plan \cite{ctsplan}.

\begin{table}[t]
\caption{NIST baseline system performance using on the 2020 CTS Challenge progress and test sets.}
\begin{center}
\begin{tabular}{|l|c|c|c|c|c|c|c|}
 \hline
\textbf{System} & \textbf{Approach} & \textbf{Training Data} & \textbf{Backend} &\textbf{Set} & \textbf{EER [\%]} & \textbf{min\_C} \\ 
 \hline\hline
 \multirow{4}{*}{NIST baseline} & \multirow{4}{*}{x-vector} & \multirow{4}{*}{CTS Superset} & \multirow{2}{*}{Cosine} & Progress & 4.37 & 0.190 \\ 
 
 &  &  &  & Test & 4.74 & 0.206 \\\cline{4-7}

 &  &  & \multirow{2}{*}{PLDA} & Progress & 4.62 & 0.221 \\
 
  &  &  &  & Test & 4.67 & 0.224 \\
 \hline
\end{tabular}
\label{tbl:results1}
\end{center}
\end{table}

Table~\ref{tbl:results1} summarizes the baseline results on the 2020 CTS Challenge progress and test sets. Note that no calibration is applied to the baseline system output. It is also worth emphasizing here that one could potentially exploit publicly available and/or proprietary data such as the VoxCeleb to further improve the performance; nevertheless, this is beyond the scope of the baseline system, and therefore not considered in this report.

\section{Acknowledgement}

Experiments and analyses were performed, in part, on the NIST Enki HPC cluser.

\section{Disclaimer}

The NIST baseline system was developed to support speaker recognition research. Comparison of systems and results against this system and its results are not to be construed or represented as endorsements of any participant's system or commercial product, or as official findings on the part of NIST or the U.S. Government. The reader of this report acknowledges that changes in the data domain and system configurations, or changes in the amount of data used to build a system, can greatly influence system performance.

Because of the above reasons, this system should not be used for commercial product testing and the results should not be used to make conclusions regarding which commercial products are best for a particular application.

\break
\bibliographystyle{IEEEtran}
\bibliography{references}
\end{document}